\documentclass[showpacs,amsmath,nofootinbib,amssymb,twocolumn,superscriptaddress]{revtex4-1}
\usepackage{array}
\usepackage{color}
\usepackage{dcolumn}
\usepackage{bm}
\usepackage{epsf}
\usepackage{graphicx}
\usepackage{amsmath} 

\newcommand{\beq}{\begin{equation}}
\newcommand{\eeq}{\end{equation}}
\newcommand{\bean}{\begin{eqnarray*}}
\newcommand{\eean}{\end{eqnarray*}\noindent}
\newcommand{\bea}{\begin{eqnarray}}
\newcommand{\eea}{\end{eqnarray}\noindent}

\begin{document}
\topmargin0in
\textheight 8.in 
\bibliographystyle{apsrev}
\title{Linearizing neutrino evolution equations including $\nu\bar{\nu}$ pairing correlations}

\author{Daavid V\"a\"an\"anen}
\email{vaananen@in2p3.fr}
\affiliation{AstroParticule et Cosmologie (APC), Universit\'e Paris Diderot - Paris 7, 10, rue Alice Domon et L\'eonie Duquet, 75205 Paris cedex 13, France}

\author{Cristina Volpe}
\email{volpe@apc.univ-paris7.fr}
\affiliation{AstroParticule et Cosmologie (APC), Universit\'e Paris Diderot - Paris 7, 10, rue Alice Domon et L\'eonie Duquet, 75205 Paris cedex 13, France}

\date{\today}


\begin{abstract}

We linearize the neutrino mean-field evolution equations describing the neutrino propagation in a background of matter and of neutrinos, using techniques from many-body microscopic approaches. The procedure leads to an eigenvalue equation that allows to identify instabilities in the evolution, associated with a change of the curvature of the neutrino energy-density surface. Our result includes all contributions from the neutrino Hamiltonian and is generalizable to linearize the equations of motion at an arbitrary point of the evolution. We then consider the extended equations that comprise the normal mean field as well as the abnormal mean field that is associated with neutrino-antineutrino pairing correlations. We first re-derive the extended neutrino Hamiltonian and show that such a Hamiltonian can be diagonalized by introducing a generalized Bogoliubov-Valatin transformation  with  quasi-particle operators that  mix neutrinos and antineutrinos. We give the eigenvalue equations that determine the energies of the quasi-particles eigenstates. Finally we derive the eigenvalue equation of the extended equations of motion, valid in the small amplitude approximation. Our results apply to an arbitrary number of neutrino families.  

\end{abstract}

\maketitle
%%%%%%%%%%%%%%%%%%%%%%%%%%%%%%%%%%%%%%%%%%%%%%%%%%%%%%%%%%%%%%%%%%%%%%%%%%%%%%%%
\section{Introduction} \label{sec:intro}
\noindent
Understanding neutrino flavor conversion in media is fascinating theoretically and crucial for observations. The history of the solar neutrino deficit problem
constitutes a reference paradigm. Davis' pioneering observations \cite{Davis:1968cp} and several decades experiments have cumulated detailed information on solar electron neutrinos.
In 1998 Super-Kamiokande has discovered the property that neutrinos modify their flavor while traveling \cite{Fukuda:1998mi}.
The SNO experiment has measured that the solar total neutrino flux is  consistent with the Standard Solar model predictions, and established the $\nu_e$ conversion into $\nu_{\mu}, \nu_{\tau}$ \cite{Ahmad:2002jz}. Finally the reactor experiment KamLAND  has identified the Large Mixing Angle solution of the solar neutrino deficit problem \cite{Eguchi:2002dm}. The ensemble of these observations establish that the high energy ($^{8}$B) electron neutrinos produced in the sun   
adiabatically convert into the other active flavors because of  the
Mikheev-Smirnov-Wolfenstein effect \cite{Wolfenstein:1977ue,Mikheev:1986gs} (see Ref.\cite{Robertson:2012ib} for a recent review).
 This is a mechanism due to the neutrino interaction with matter, which has become  a reference in our knowledge of how neutrinos can modify their flavor in media. 

A variety of novel phenomena has been unravelled by the study of neutrinos in astrophysical and cosmological environments, as e.g. core-collapse supernovae. In this context, the inclusion of the neutrino 
self-interaction has shown new features such as
collective effects (see \cite{Duan:2010bg} for a review), that can be understood in terms of flavor \cite{Duan:2007mv} or gyroscopic pendulum \cite{Hannestad:2006nj}, as an MSW-like solution in the comoving frame \cite{Raffelt:2007cb} or a magnetic resonance \cite{Galais:2011gh}.  The presence of (front and reverse) shocks in such explosive environments engenders multiple MSW resonances (see  \cite{Duan:2009cd} for a review). 
The investigation of the phenomena occurring is certainly necessary in view of the possible implications for supernova physics and observations, and to establish the definite interplay with unknown neutrino properties, such as the neutrino mass hierarchy \cite{Serpico:2011ir,Gava:2009pj} and leptonic CP violation  \cite{Balantekin:2007es,Gava:2008rp}. While numerous aspects are understood, a comprehensive understanding of how neutrinos modify their flavor in these explosive media requires further investigations.

Recently, the connection is being developed between formal aspects of the description of neutrino flavor conversion in media and of other many-body systems like atomic nuclei and condensed matter. 
Establishing these links is certainly fascinating theoretically.  Moreover it can give novel paths to go beyond the currently adopted description, or test the validity of approximations that are widely employed.
In Ref.\cite{Balantekin:2006tg} the $\nu$-$\nu$ Hamiltonian is reformulated using an algebraic approach. Corrections to the commonly used mean-field Hamiltonian are obtained using the path-state integral approach.
This algebraic approach is pursued in Ref.\cite{Pehlivan:2011hp} where the many-body Hamiltonian (without the neutrino-matter interaction term) is put in connection with the Bardeen-Cooper-Schrieffer (BCS) theory of superconductivity. A Random-Phase-approximation (RPA) version of the equations is given. The spectral-split phenomenon due to the $\nu$-$\nu$
interaction is understood as the transition from the quasi-particle to the particle degrees of freedom, with the Lagrange multipliers being related to the split energy. In Ref.\cite{Volpe:2013uxl} the Bogoliubov-Born-Green-Kirkwood-Yvon (BBGKY) theoretical framework, that is widely used in the study of many-body systems, is adopted to re-derive the mean-field equations describing 
the neutrino propagation in a background of matter and of neutrinos. Besides giving a rigorous derivation of such equations, such an approach allows to go beyond the mean-field approximation to include contributions from the two-body density matrix, like neutrino-antineutrino pairing correlations. Note that these contributions have been neglected  so far (see e.g. \cite{Sigl:1992fn,Cardall:2007zw}). However,  as demonstrated in Ref.\cite{Volpe:2013uxl} such corrections already appear at the mean-field level. Since the equations are non-linear, their impact on supernova physics and on neutrino flavor conversion in such environments might be significant, and deserves investigation.
Linearizing evolution equations around a stationary solution is another general procedure, commonly employed in the context of microscopic many-body approaches. In fact, such
methods give eigenvalue equations that identify the presence of collective modes, such as giant resonances in atomic nuclei or phonons in metallic clusters. These are also used to detect instabilities that typically occur when correlations become too large and induce a change of the energy-density curvature around the starting point of the linearization procedure. 

In the specific context of core-collapse supernova neutrinos, 
demanding numerical calculations are required to implement realistically the neutrino-neutrino interaction, or the dynamical features associated with the explosion. A detailed treatment of the transition between the region where neutrinos are trapped, and the one where they start free streaming (the neutrinosphere) will require still some time, in order to properly take into account neutrino rescattering for example \cite{Cherry:2012zw} or two-body density matrix contributions, such as $\nu$-$\bar{\nu}$ pairing correlations \cite{Volpe:2013uxl}. The linearization procedure might bring some insight on these complex cases, without going through full simulations.
In situations encountered in many-body systems,  typically stable collective small amplitude modes are searched for. In the neutrino context such an approach appears to be of particular use for the study of unstable modes.
In fact, the presence of flavor instabilities is a characteristic feature of the neutrino evolution in a medium, such as a core-collapse supernova, where the non-linear neutrino self-interaction is important.
In Ref.\cite{Hannestad:2006nj} it has been pointed out that the bipolar oscillations in the single-angle approximation is associated with an instability triggered by the presence of the vacuum mixings. Using the matter basis within two-neutrino flavors, Ref.\cite{Galais:2011jh} has given an analytical condition to identify the occurrence of such an instability. 
Such a condition is consistent with the one derived heuristically in Ref.\cite{Duan:2010bf}.
Multi-angle instabilities are investigated in Ref.\cite{Sawyer:2008zs}.
Besides,   \cite{Sawyer:2008zs} has first pointed out the use of linearization methods
to investigate such instabilities. Ref.\cite{Banerjee:2011fj} has given an eigenvalue equation (for two-flavors) and studied numerically their appearance,  both in the simplified (single-angle) angular treatment of the neutrino-neutrino interaction term, and in sophisticated multi-angle calculations. Schematic (box-like) neutrino fluxes typical of the supernova case are used. A linearized flavor stability analysis is performed in \cite{Sarikas:2011am,Saviano:2012yh} in realistic cases, i.e. in order to study the suppression of collective effects during the accretion phase of various iron-core collapse supernova progenitors.
Ref.\cite{Sarikas:2012ad} has used the same method to show the presence of spurious instabilities when the number of angular bins used in multi-angle calculations is not large enough. The presence of azimuthal angle instabilities is identified in Ref.\cite{Raffelt:2013rqa}.

The present work is based on the extended neutrino evolution equations, including neutrino-antineutrino pairing correlations, derived in 
Ref.\cite{Volpe:2013uxl} using the BBGKY hierarchy.
We first apply a linearization method known in the context of many-body approaches to the neutrino evolution equations. We derive a general eigenvalue equation that implements the vacuum, the matter and the neutrino-neutrino interaction terms of the Hamiltonian. We introduce the stability matrix associated with the neutrino  energy density surface. Then, we consider the extended evolution equations containing contributions from the abnormal mean field associated with neutrino-antineutrino pairing correlations. As for Ref.\cite{Volpe:2013uxl} here we focus on the formal aspects while simulations including such terms are beyond the scope of the present work. 
 Two aspects of the case with $\nu$-$\bar{\nu}$ pairing are studied: the static solution of the generalized Hamiltonian including pairing correlations and a linearization of the corresponding evolution equations. For the first, a  re-derivation of the Hamiltonian with $\nu\bar{\nu}$ correlations is given in the mean-field approximation. 
Using a generalized Bogoliubov-Valatin transformation that mixes neutrinos and their $CP$-conjugates, we demonstrate that the eigenstates of the system described by the generalized Hamiltonian are independent quasi-particle states. The eigenvalue equation to determine the eigen-energies of the ground- and of the excited quasi-particle states is also given. 
Finally we present the linearized version of  the extended equations of motion obtained in \cite{Volpe:2013uxl}, which can be used to study both stable and unstable small amplitude collective modes due to $\nu$-$\bar{\nu}$ pairing.

The manuscript is organised as follows. In Section I we introduce the theoretical framework and present the extended  mean-field neutrino evolution equations describing neutrino propagation in an environment. These include the neutrino mixings, the neutrino interaction with matter, with neutrinos and neutrino-antineutrino correlations. In Section II we present our linearization procedure. We apply it  to the case where only the normal mean field is included to obtain the eigenvalue equations and the stability matrix. Section III focuses on the static case of the extended Hamiltonian, its diagonalization is performed by introducing quasi-particle states and the equations to determine the eigen-energies are derived. 
Section IV presents the linearized version of the extended evolution equations. A discussion and our conclusions are contained in Section V.
 
 \section{Neutrino mean-field evolution equations in media with $\nu$-$\bar{\nu}$ pairing}
\noindent
Having in mind astrophysical and cosmological applications, we consider a system of neutrinos and antineutrinos propagating in a medium composed of matter, neutrinos and antineutrinos. 
The corresponding Hamiltonian in the flavor basis is 
 \beq\label{e:Hf}
H^f = UH_{vac}U^{\dagger} + H_{int} 
\eeq
where the vacuum term is $H_{vac} = diag(E_{m_i})$ with eigenenergies $E_{m_i}$ for the propagation eigenstates with mass $m_i$ ($i=1, 2, 3$). The second term $H_{int}$ corresponds to the interaction between a neutrino and any other particle of the background.  The unitary matrix  $U$ is the Pontecorvo-Maki-Nakagawa-Sakata (PMNS) matrix  \cite{Maki:1962mu}  that relates the interaction (flavor) to the propagation (mass) basis through $  |\nu_{\alpha} \rangle= \sum_{i} U^*_{\alpha i} |\nu_i \rangle$. In three flavors, such a matrix depends upon three  measured mixing angles, one Dirac and two Majorana unknown phases \cite{PDG2012}.

Within a density matrix formalism, the information on the flavor evolution is encoded in the density matrix that reads, for three flavors,
 \beq\label{e:rho}
\rho_{\nu} = \left(
\begin{array}{ccc} \langle a^{\dagger}_{\nu_{\alpha},i} a_{\nu_{\alpha},i} \rangle &  \langle a^{\dagger}_{\nu_{\beta},j} a_{\nu_{\alpha},i} \rangle  
&  \langle a^{\dagger}_{\nu_{\gamma},k} a_{\nu_{\alpha},i} \rangle \\
\langle a^{\dagger}_{\nu_{\alpha},i} a_{\nu_{\beta},j} \rangle
& \langle a^{\dagger}_{\nu_{\beta},j} a_{\nu_{\beta},j} \rangle  & \langle a^{\dagger}_{\nu_{\gamma},k} a_{\nu_{\beta},j} \rangle \\ 
 \langle a^{\dagger}_{\nu_{\alpha},i} a_{\nu_{\gamma},k} \rangle & \langle a^{\dagger}_{\nu_{\beta},j} a_{\nu_{\gamma},k} \rangle  & \langle a^{\dagger}_{\nu_{\gamma},k} a_{\nu_{\gamma},k} \rangle 
\end{array}
 \right).
\eeq
The $a_{\nu_{\alpha},i}$ and $a^{\dagger}_{\nu_{\alpha},i}$ are the annihilation and the creation operators for a neutrino of flavor $\alpha$ 
 in the quantum state identified by the single-particle label $i$, that indicates for example momentum $\vec{p}$, or helicity $h$. 
The expectation values  in Eq.(\ref{e:rho})  are performed over the full many-body density matrix which, in general, can be associated to pure or to mixed states. The diagonal elements of $\rho_{\nu}$  correspond to the neutrino occupation numbers, with $N_{\nu_{\alpha}} = \sum_i \langle a^{\dagger}_{\nu_{\alpha},i} a_{\nu_{\alpha},i} \rangle $ being the total occupation number for a neutrino of $\alpha$ flavor. The off-diagonal (or decoherent) terms encode the neutrino mixings. 
Note that a "matrix of densities" generalizing the usual occupation numbers is commonly used in the literature (see e.g. \cite{Raffelt:1992uj,Sigl:1992fn}). 
The density matrix $\rho_{\nu}$ is easily extended to an arbitrary number of families,
to account for the presence of both sterile and active neutrinos. 
A definition analogous to (\ref{e:rho}) is introduced for antineutrinos, whose explicit expression is
\beq\label{e:rhoanu}
\bar{\rho}_{\nu} = \left(
\begin{array}{ccc} \langle b^{\dagger}_{\nu_{\alpha},i} b_{\nu_{\alpha},i} \rangle &  \langle b^{\dagger}_{\nu_{\beta},j} b_{\nu_{\alpha},i} \rangle  
&  \langle b^{\dagger}_{\nu_{\gamma},k} b_{\nu_{\alpha},i} \rangle \\
\langle b^{\dagger}_{\nu_{\alpha},i} b_{\nu_{\beta},j} \rangle
& \langle b^{\dagger}_{\nu_{\beta,j}} b_{\nu_{\beta},j} \rangle  & \langle b^{\dagger}_{\nu_{\gamma},k} b_{\nu_{\beta},j} \rangle \\ 
 \langle b^{\dagger}_{\nu_{\alpha},i} b_{\nu_{\gamma},k} \rangle & \langle b^{\dagger}_{\nu_{\beta},j} b_{\nu_{\gamma},k} \rangle  & \langle b^{\dagger}_{\nu_{\gamma},k} b_{\nu_{\gamma},k} \rangle
 \end{array}
\right).
\eeq
where $b$ and $b^{\dagger}$ the annihilation and creation operators for antineutrinos.  
The creation and annihilation particle and antiparticle operators satisfy the usual canonical commutation rules\footnote{Note that for relativistic neutrinos an approximate Fock space can be built (see \cite{Giunti:1991cb} for a discussion).} 
\beq\label{e:commutators1}
\{ a(\vec{p},h), a^{\dagger}(\vec{p}\,',h') \} 
 = (2 \pi)^3 2 E_{p}\delta^3(\vec{p} - \vec{p}\,')\delta_{hh'}
\eeq 
and similarly for the antiparticle operators. All other anticommutators vanish.
The single-particle states associated with neutrino mass eigenstates are
\beq\label{e:sp}
| m \rangle = a^{\dagger}_m | \rangle~Ê~Ê| m \rangle = b^{\dagger}_m | \rangle
\eeq
with $| \rangle$ being the vacuum state defined by $a_m | \rangle = 0$ and $b_m | \rangle = 0$.

In Ref.\cite{Volpe:2013uxl} we have derived extended mean-field equations to describe a neutrino, or an antineutrino, evolving in a matter and in a(n) (anti)neutrino background using the BBGKY hierarchy \cite{Born-Green,Yvon,Kirkwood,Bogoliubov}. This corresponds to an (unclosed) set of coupled first-order integro-differential equations for the reduced density matrices, which is equivalent to determining the exact evolution of the full many-body density matrix. Note that while BBGKY is often formulated in the density matrix formalism, it is formally equivalent to an (infinite) hierarchy of equations for the n-point Green functions in the equal-time limit. Such a hierarchy gives a framework to re-derive the usually employed mean-field and Boltzmann approximations, but also to go beyond them in a consistent way. In particular, if one includes neutrino-antineutrino pairing correlations, one arrives at extended mean-field equations that can be cast in a compact matrix form (see \cite{Volpe:2013uxl} for the details of the derivation). In the flavor basis these read 
\begin{equation}\label{e:ex}
i \dot{{\cal R}} = [ {\cal H},{\cal R}],
\end{equation}
with  ${\cal R}$ the generalized density:
\begin{equation}\label{e:rex}
{\cal R} = \left(
\begin{array}{cc}   
 \rho &  \kappa  \\
\kappa^{\dagger} & 1 - \bar{\rho}^* \end{array}
\right).
\end{equation}
${\cal R}$ depends on the normal densities $ \rho$ for $\nu$ Eq. (\ref{e:rho}) and  $\bar{\rho}$ for $\bar{\nu}$, on the abnormal density:
\beq\label{e:kmat}
\kappa_{\nu} = \left(
\begin{array}{ccc} \langle b_{\nu_{\alpha},i} a_{\nu_{\alpha},i} \rangle &  \langle b_{\nu_{\beta},j} a_{\nu_{\alpha},i} \rangle  
&  \langle b_{\nu_{\gamma},k} a_{\nu_{\alpha},i} \rangle \\
\langle b_{\nu_{\alpha},i} a_{\nu_{\beta},j} \rangle
& \langle b_{\nu_{\beta,j}} a_{\nu_{\beta},j} \rangle  & \langle b_{\nu_{\gamma},k} a_{\nu_{\beta},j} \rangle \\ 
 \langle b_{\nu_{\alpha},i} a_{\nu_{\gamma},k} \rangle & \langle b_{\nu_{\beta},j} a_{\nu_{\gamma},k} \rangle  & \langle b_{\nu_{\gamma},k} a_{\nu_{\gamma},k} \rangle
 \end{array}
\right),
\eeq
as well as on its complex conjugate 
\begin{equation}\label{e:kmat*}
\kappa^*_{\nu}  =  \left(
\begin{array}{ccc} \langle a^{\dagger}_{\nu_{\alpha},i} b^{\dagger}_{\nu_{\alpha},i} \rangle &  \langle a^{\dagger}_{\nu_{\beta},j} b^{\dagger}_{\nu_{\alpha},i} \rangle  
&  \langle a^{\dagger}_{\nu_{\gamma},k} b^{\dagger}_{\nu_{\alpha},i} \rangle \\
\langle a^{\dagger}_{\nu_{\alpha},i} b^{\dagger}_{\nu_{\beta},j} \rangle
& \langle a^{\dagger}_{\nu_{\beta},j} b^{\dagger}_{\nu_{\beta},j} \rangle  & \langle a^{\dagger}_{\nu_{\gamma},k} b^{\dagger}_{\nu_{\beta},j} \rangle \\ 
 \langle a^{\dagger}_{\nu_{\alpha},i} b^{\dagger}_{\nu_{\gamma},k} \rangle & \langle a^{\dagger}_{\nu_{\beta},j} b^{\dagger}_{\nu_{\gamma},k} \rangle  & \langle a^{\dagger}_{\nu_{\gamma},k} b^{\dagger}_{\nu_{\gamma},k} \rangle 
\end{array}
 \right).
\end{equation}
These encode the $\nu$-$\bar{\nu}$  pairing correlations. The generalized Hamiltonian ${\cal H}$ in Eq.(\ref{e:ex}) is given by
\begin{equation}\label{e:hex}
{\cal H} = \left(
\begin{array}{cc}   
h & \Delta \\
\Delta^{\dagger} & -\bar{h}^* \end{array}
\right).
\end{equation}
It comprises the mean-field Hamiltonian $h$ for neutrinos, $\bar{h}$ for antineutrinos  as well as the abnormal (or pairing) mean field $\Delta$.  Its general expression is 
\beq\label{e:abpair}
\Delta_{ik} = \sum_{jl} v_{(ik,jl)} \kappa_{jl}.
\eeq
Its complex conjugate $\Delta^*$ depends on $ \kappa^*_{jl}$ and involves a sum over the initial single-particle states, instead of over the final single-particle states. 
The quantity :
 \beq\label{e:matel}
 v_{(im,jn)} = \langle im | H_{int} | jn \rangle 
 \eeq
 are the matrix elements associated with the interaction $H_{int}$ Eq.(\ref{e:Hf}) which involve   $jn $ and $im$ single-particle states Eq.(\ref{e:sp}) 
 associated with the incoming and outgoing particles. Therefore from Eqs.(\ref{e:ex}-\ref{e:kmat*}) one can see that to determine the evolution of the system, within the extended mean field, one has to identify the solution of a coupled system of equations for $\rho_{\nu}$, for $\bar{\rho}_{\nu}$ and for  their two-body correlators 
$\kappa_{\nu}$ and $\kappa^*_{\nu}$.
 The explicit expression of the pairing mean fields $\Delta$ and $\Delta^*$ depend on the assumptions of homogeneity and isotropy made on the background. In particular, if homogeneity is assumed, $\kappa$ involves pairs of neutrinos and antineutrinos with opposite momentum, as discussed in \cite{Volpe:2013uxl}. In the present manuscript we will only need the general expressions of the pairing mean fields (the explicit expression in cartesian or polar coordinates can be found in \cite{Volpe:2013uxl}).

In absence of $\nu$-$\bar{\nu}$ pairing,  Eqs.(\ref{e:ex}) reduce to the mean-field evolution equation:
\beq\label{e:mf}
i \dot{\rho}  =  [h(\rho),\rho ] 
\eeq
with the normal mean field acting on a neutrino given by
\begin{eqnarray}\label{e:mfnu}
 h(\rho) &= &UH_mU^{\dagger} +  H_{mat}  \\
 && + \sqrt{2} G_F \sum_{\underline{\alpha}} \int {{d^3 \vec{p}} \over{(2 \pi)^3 }} (\rho_{\underline{\alpha},p} - \bar{\rho}^*_{\underline{\alpha},p}) \left( 1- \hat{\vec{p}} \cdot \hat{\vec{k}} \nonumber
\right) .
\end{eqnarray}
The term accounting for neutrino interaction with matter is $H_{mat} = diag(\sqrt{2} G_F n_e, 0, 0)$, with $G_F $ being the Fermi coupling constant and $n_e$ the electron number density. The  $\underline{\alpha}$ refers to a neutrino that is initially born in the $\alpha$ flavor since one has to sum over all the flavors present in the background. 
In deriving Eq.(\ref{e:mfnu}) the electron background is assumed homogenous and isotropic, while the (anti)neutrino background, homogenous and anisotropic. 
The anisotropy introduces the angular term depending on $\hat{\vec{p}} \cdot \hat{\vec{k}}$ defined as
\beq\label{z:}
\hat{\vec{p}} \cdot \hat{\vec{k}} = {{\vec{p} \cdot \vec{k}} \over{ |\vec{p} |  \cdot |\vec{k} |  }}.
\eeq

In case an antineutrino is traveling instead of a neutrino, an equation analogous to Eq.(\ref{e:mf}) holds 
\beq\label{e:mfa}
i \dot{\bar{\rho}} =  [ \bar{h}(\bar{\rho}) ,\bar{\rho} ] 
\eeq
with
the corresponding normal mean field\footnote{Note that, our definition is $\bar{\rho}= \langle b^{\dagger}_{\nu_{\alpha},i} b_{\nu_{\alpha},i} \rangle$ (see Eq.(\ref{e:rhoanu})), so that antineutrinos do not transform the same way as neutrinos under $U$.}  given by 
\begin{eqnarray}\label{e:mfanu}
 \bar{h}(\bar{\rho}) &= &U^*H_mU^{T} - H_{mat}   \\
 && - \sqrt{2} G_F \sum_{\underline{\alpha}} \int {{d^3 \vec{p}} \over{(2 \pi)^3 }} (\rho^*_{\underline{\alpha},p} - \bar{\rho}_{\underline{\alpha},p}) \left( 1- \hat{\vec{p}} \cdot \hat{\vec{k}} \nonumber
\right). 
\end{eqnarray}
In case of an isotropic background, the contribution to the integral coming from the momentum scalar products $\hat{\vec{p}} \cdot \hat{\vec{k}}$ vanishes. In this case only the contribution coming from the density matrices remains, giving a term analogous to $H_{mat}$, but with the (anti)neutrino number densities replacing the electron one. The derivation of the mean-field equations (\ref{e:mf}-\ref{e:mfanu}) can be found in \cite{Volpe:2013uxl}, and in \cite{Sigl:1992fn,Qian:1994wh,Balantekin:2006tg} (using different methods).

In component form, the extended equations (\ref{e:ex}-\ref{e:hex}) in presence of $\nu\bar{\nu}$ pairing are 
\begin{equation} \label{e:tddm}
\left\{ 
\begin{array}{lcl} 
%\begin{eqnarray}  
i \dot{\rho}_{ij} (1) & = & [h(1),\rho(1)]_{ij} + \sum_m (\Delta_{im} \kappa^*_{jm} - \kappa_{im} \Delta^*_{jm}  ) \\ 
i \dot{\bar{\rho}}_{kl} (2) & = & [\bar{h}(2),\bar{\rho}(2)]_{kl} + \sum_m (\Delta_{mk} \kappa^*_{ml} - \kappa_{mk} \Delta^*_{ml}  ) \\ 
i \dot{\kappa}_{ik} & = & \sum_m (h_{im}(1) \kappa_{mk} + h_{km}(2) \kappa_{im}) + \Delta_{ik}   \\ 
 &  & - \sum_m (\rho_{im} (1) \Delta_{mk} + \bar{\rho}_{km} (2) \Delta_{im} )
%  \end{eqnarray} \right .
\end{array} \right. ,
\end{equation}
where here the indices $i,j$ stand for $\nu_{\alpha}, \nu_{\beta}$; $k, l$ for  $\bar{\nu}_{\alpha}, \bar{\nu}_{\beta}$  with $\alpha, \beta $ that vary over  the different electron, muon and tau flavor states.
For the sake of clarity, Eqs.(\ref{e:tddm}) we have introduced an explicit dependence of the quantities on particles of type 1 corresponding to neutrinos, and on particles
of type 2 referring to antineutrinos since $\kappa$ correlates the two kind of particles. 

\section{Linearization procedure and its application to the mean-field equations}
\noindent
Theoretical and phenomenological studies of neutrino flavor conversion in astrophysical environments typically solve Eqs.(\ref{e:mf})-(\ref{e:mfanu}) in either schematic models, or realistic cases with the goal of underpinning the physical mechanisms and/or making reliable predictions for supernova observables. Linearizing the equations of motion is a standard procedure in the study of the many-body systems and leads to microscopic approaches known as the Random-Phase-Approximation (RPA), or the Quasi-Particle Random-Phase-Approximation (QRPA) \cite{Ring,Tohyama:2004ed}. Here we first describe such a method and then apply it to the neutrino case when the mean-field approximation is made.

\subsection{Linearization procedure}
\noindent
Let us consider a system described by a mean-field Hamiltonian $h$ and a one-body density matrix $\rho$. In the mean-field approximation the evolution equation is
\beq\label{e:tdhf}
i \dot{\rho}  =  [h(\rho),\rho ]. 
\eeq
Let us consider the case that the system is in a stationary solution at a given time $t_0$ 
$\rho^{0}=\rho (t=t_0)$, then
\beq\label{e:hf}
[h^{0},\rho^{0}]= 0
\eeq
where $h^{0}=h(\rho^{0})$. This implies that there is a basis in which both
$\rho^{0}$ and the mean-field Hamiltonian $h^{0}$ are diagonal. 
Note that the energy variation in this basis  
is such that $\delta E = E(\rho^{0} + \delta \rho) - E(\rho^{0}) = 0$,
 with $E=tr(h \rho)$. Therefore Eq.(\ref{e:hf}) identifies the stationary solution that minimizes the energy of the system. 

We now consider a small variation $\delta \rho(t)$ of the density around $\rho^{0}$:
\beq\label{e:rho1}
\delta \rho = \rho_0 + \delta \rho(t) = \rho^{0} +  \rho'e^{-i\omega t} +  \rho'^{\dagger} e^{i\omega^* t}.  
\eeq
In the specific case where the initial state is a pure state, i.e. $\rho^2=\rho$ it is easy to show that such a variation $\delta \rho$  can only have nonzero contributions if $\rho'_{im}=\langle a^{\dagger}_m a_i \rangle $\footnote{In the context of many-body approaches,  these contributions are usually referred to as of particle-hole type, $ \rho^{ph}$\cite{Ring}. 
The basis diagonalizing the mean field $h$ is usually called the Hartree-Fock basis.} where $m$ and $i$ are unoccupied and occupied single-particle states respectively.  
This is for example a good approximation for the case for atomic nuclei whose ground states are well described by Slater determinants where the nucleons build up a Fermi sea of occupied single-particle states. 
In the case of mixed states as for neutrinos produced in dense stellar regions one can associate $\delta \rho$ with the decoherent off-diagonal contributions
that are present because of the mixings.
Physically speaking, the $\omega$ frequency in Eq.(\ref{e:rho1}) represents the frequencies of the small amplitude excitation modes of the system around $\rho^{0}$. 

Usually, one introduces an external field $F(t)$ that drives the system out of the equilibrium solution $\rho^{0}$ and induces small amplitude variations
around such a solution (see e.g. \cite{Ring}). For our cases of interest,  
we make the hypothesis that at $t=t_0$ the density matrix  is "stationary" to a good approximation 
if the time-dependence of Hamiltonian is weak so that  the system stays in such a solution for some time.
Then the time-dependence of the Hamiltonian drives the system out of equilibrium\footnote{As we we will discuss an example of this kind is given by the transition between the synchronization and the bipolar regimes produced by the neutrino-neutrino interaction in the context of core-collapse supernovae (see e.g. \cite{Duan:2007mv,Hannestad:2006nj,Galais:2011jh}).}. 
In order to study the behaviour of the system in this small amplitude approximation, one can linearize the neutrino evolution equations (\ref{e:tdhf}) around the solution $\rho^{0}$ satisfying Eq.(\ref{e:hf}). The development of the mean-field Hamiltonian around this solution gives
\beq\label{e:linh}
h(\rho) =  h^{0} + {\delta h \over {\delta \rho}}\Big|_{\rho^{0}}\delta \rho + \ldots 
\eeq
Implementing the small amplitude variation Eq.(\ref{e:rho1}) and
retaining only the lowest order contributions, one gets 
\begin{eqnarray}\label{e:linh2}
\omega \rho' e^{-i\omega t} - \omega^* \rho'^{*}e^{i\omega^* t} & = & [h^{0},\rho^{0} + \delta \rho] \\
&+& \left[ {\delta h \over {\delta \rho}} \delta \rho,\rho^{0} \right] \nonumber .
\end{eqnarray}
From Eq.(\ref{e:hf}) one obtains for the first commutator on the {\it r.h.s.} of Eq.(\ref{e:linh2}) 
\beq\label{e:c1}
[h^{0},\delta \rho]_{ij} = (\tilde{k}_i - \tilde{k}_j) \delta \rho_{ij},
\eeq
where $\tilde{k}_i$ are the energy eigenvalues associated with the 
single-particle states that diagonalize the Hamiltonian $h^{0}$.
The second commutator on the {\it r.h.s.} of Eq.(\ref{e:linh2}) gives 
\begin{align}\label{e:c2}
\left[{\delta h \over {\delta \rho}} \delta \rho,\rho^{0} \right]_{ij}  & = \sum_k \left({\delta h \over {\delta \rho}}\Big|_{\rho^{0}} \delta \rho \right)_{ik} \rho^{0}_{kj} \delta_{kj} \\ \nonumber
& \quad - \rho^{0}_{ik}  \delta_{ik} \left({\delta h \over {\delta \rho}}\Big|_{\rho^{0}} \delta \rho \right)_{kj} , 
\end{align}
where we have used the fact that  we are in the basis in which $\rho$ has only nonzero diagonal elements.

By introducing  Eqs.(\ref{e:hf}) and (\ref{e:c1}-\ref{e:c2})  in (\ref{e:linh2}) a general eigenvalue equation is obtained, valid in the small amplitude approximation \cite{Ring} :
\begin{align}\label{e:tda}
\omega \rho'_{ij}  &=  \sum_{kl}
 \left[ (\tilde{k}_k - \tilde{k}_l)\delta_{ki} \delta_{jl} \rho'_{kl}  + (\rho^{0}_{j} - \rho^{0}_{i}) 
{\delta h_{ij} \over {\delta \rho'_{kl}}} \Big|_{\rho^{0}} \rho'_{kl}  \right] . 
\end{align}
A similar equation holds for $\omega^* \rho'^{*}_{ij}$, instead of $\omega \rho'_{ij}$.

To gather further insight in the behaviour of a system around a stationary solution of the Hamiltonian, one can 
also consider  developing the energy density of the system in the small amplitude approximation
\begin{equation}\label{e:encurv}
E[\rho] = E[\rho^0] + {\delta E \over{\delta \rho}}\Big|_{\rho^{0}} \delta \rho + {\delta^2 E \over{\delta \rho^2}}\Big|_{ \rho^{0}}  \delta \rho^2 + \ldots
\end{equation}
where for the stationarity one has ${\delta E /{\delta \rho}}|_{\rho^{0} }= 0$. 
The second derivative of the energy density defines the the energy-density curvature around $ \rho^{0}$. One can show that such a term can be associated to a {\it stability matrix} that is positive definite in presence of stable small amplitude modes\footnote{Note that this argument can be generalized to a multi-variable energy density by considering its Hessian.}. The presence of complex eigenvalues of the stability matrix signals a change in the curvature in presence of an instability \cite{Ring}. Therefore, in order to identify small amplitude modes of the system requires solving the eigenvalue equations (\ref{e:tda}) obtained by linearizing the equations of motion or, equivalently, identify the eigenvalues of the stability matrix. Under some extra assumptions on the density matrix variations, that depend on the specific system under consideration, one can also construct the stability matrix directly from Eq.(\ref{e:encurv}). The latter procedure is for example of use in the case of atomic nuclei where $\delta \rho$, as $\rho$, keeps being a projector on the single-particle occupied states\footnote{In other words, $\rho$ is associated to a Slater determinant.}. In the following we will use the linearization procedure described above. In the application to the mean-field case we will also identify the stability matrix. 

\subsection{General linearized eigenvalue equations for neutrinos in media}
\noindent
We now apply the procedure just described to our system of neutrinos and antineutrinos propagating in a medium. As seen from 
Eqs.(\ref{e:mf}-\ref{e:mfanu}), the mean fields $h$ and $\bar{h}$ are a function of the two density matrices  $\rho$ and $\bar{\rho}$. At $t=t_0$ one has 
\begin{equation}\label{e:h0}
h^{0}=h(\rho^{0},\bar{\rho}^{0}),
\end{equation}
 and 
 \begin{equation}\label{e:hbar0}
\bar{h}^{0}=\bar{h}(\rho^{0},\bar{\rho}^{0}). 
\end{equation}
Therefore relations (\ref{e:linh}-\ref{e:tda}) have to be generalized to include the particle and antiparticle degrees of freedom. In particular, one has
the following variations 
\begin{align}\label{e:dh}
\delta h = {\delta h(\rho,\bar{\rho}) \over {\delta \rho}}\Big|_{(\rho^{0},\bar{\rho}^{0})}\delta \rho + {\delta h(\rho,\bar{\rho}) \over {\delta \bar{\rho}}}\Big|_{(\rho^{0},\bar{\rho}^{0})}\delta \bar{\rho}
\end{align}
\begin{align}\label{e:dhb}
\delta \bar{h} = {\delta \bar{h}(\rho,\bar{\rho}) \over {\delta \rho}}\Big|_{(\rho^{0},\bar{\rho}^{0})}\delta \rho + {\delta \bar{h}(\rho,\bar{\rho}) \over {\delta \bar{\rho}}}\Big|_{(\rho^{0},\bar{\rho}^{0})}\delta \bar{\rho}.
\end{align}
This adds an extra commutator to Eq.(\ref{e:linh2}) (similarly in the linearized equations for antineutrinos). 
For our system of neutrinos and antineutrinos, from Eq.(\ref{e:tda}) two coupled eigenvalue equations  are derived\footnote{Equation (\ref{e:tda}) written for particle-hole ($\rho^{ph}$) and hole-particle ($\rho^{hp}$) contributions gives the RPA equations. In this case the derivatives of $h$ with respect to $\delta \rho$ have two contributions and give rise to the so called ph-ph interaction terms \cite{Ring}.}
\begin{align}\label{e:tdanu}
\omega \rho'_{ij}  &=  \sum_{k l}
\Big[ (\tilde{k}_k - \tilde{k}_l)\delta_{ki} \delta_{jl} \rho'_{kl} \\  \nonumber
& \quad +  (\rho^{0}_{j} - \rho^{0}_{i})   \Big(
{\delta h_{ij} \over {\delta \rho'_{kl}}}\Big|_{(\rho^{0},\bar{\rho}^{0})} \rho'_{kl} +
 {\delta h_{ij} \over {\delta \bar{\rho}'_{kl}}}\Big|_{(\rho^{0},\bar{\rho}^{0})}\bar{\rho}'_{kl} \Big) \Big] \\  \nonumber
\omega  \bar{\rho}'_{ij}  &=  \sum_{k l}
\Big[ (\tilde{ \bar{k}}_l - \tilde{ \bar{k}}_k)\delta_{kj} \delta_{il}  \bar{\rho}'_{kl} \\  \nonumber
& \quad +  ( \bar{\rho}^{0}_{j} -  \bar{\rho}^{0}_{i})   \Big(
{\delta  \bar{h}_{ij} \over {\delta  \bar{\rho}'_{kl}}}\Big|_{(\rho^{0},\bar{\rho}^{0})} \bar{\rho}'_{kl} +
 {\delta  \bar{h}_{ij} \over {\delta \rho'_{kl}}}\Big|_{(\rho^{0},\bar{\rho}^{0})}\rho'_{kl}  \Big) \Big]. 
\end{align}
Similar equations hold for $\omega^* \rho'^{*}_{ij}  $ and $\omega^*  \bar{\rho}'^{*}_{ij} $.

Eq.(\ref{e:tdanu}) can be cast in a compact matrix form :
\begin{align}\label{e:lin1}
\left(
\begin{array}{ll}
A & B \\
\bar{B} & \bar{A} \\
\end{array} 
\right) \left(
\begin{array}{l}
\rho'  \\
\bar{\rho}' \\  
\end{array}
\right) 
 & = \omega
\left(
\begin{array}{l}
\rho'  \\
\bar{\rho}' \\  
\end{array}
\right), 
\end{align}
with 
\begin{align}\label{e:RPA}
A_{ij,kl} & = \tilde{k}_i - \tilde{k}_j + (\rho^{0}_{j} - \rho^{0}_{i}) {\delta h_{ij}\over \delta \rho_{kl}}\Big|_{(\rho^{0},\bar{\rho}^{0})}  \nonumber \\
B_{ij,kl}  & =  (\rho^{0}_{j} - \rho^{0}_{i}) {\delta h_{ij}\over \delta \bar{\rho}'_{kl}}\Big|_{(\rho^{0},\bar{\rho}^{0})}  \nonumber \\
\bar{A}_{ij,kl} & = (\tilde{\bar{k}}_j - \tilde{\bar{k}}_i) +  ( \bar{\rho}^{0}_{i} -  \bar{\rho}^{0}_{j})  
{\delta  \bar{h}_{ji} \over {\delta  \bar{\rho}'_{kl}}}\Big|_{(\rho^{0},\bar{\rho}^{0})} \nonumber \\
\bar{B}_{ij,kl} & =  ( \bar{\rho}^{0}_{i} -  \bar{\rho}^{0}_{j}) {\delta  \bar{h}_{ji} \over {\delta \rho'_{kl}}}\Big|_{(\rho^{0},\bar{\rho}^{0})}.   \nonumber \\
 \end{align}
Note that the matrix form (\ref{e:lin1}) involve the components $\rho'_{ij}$ and $\bar{\rho}'_{ji}$.
For each solution of Eqs.(\ref{e:RPA}) $(\rho^*, \bar{\rho}^*) $ are also a solution of the eigenvalue equations corresponding to $\omega^*$.
The matrix appearing on ${\it l.h.s. } $ of Eq.(\ref{e:lin1}) is the {\it stability matrix}:
\begin{align}\label{e:Smat}
S & =
\left(
\begin{array}{ll}
A & B \\
\bar{B} & \bar{A} \\
\end{array} 
\right) . 
\end{align}

While the results just derived are valid for any system of neutrinos and antineutrinos described by a mean-field equation, we now make them specific to the case of 
core-collapse supernova neutrinos with the mean field given by Eqs.(\ref{e:mf}-\ref{e:mfanu}). 
Let us take as initial time the neutrinosphere where neutrinos start free-streaming. 
In this region, the large matter density suppresses the neutrino mixing angles, so that the flavor and matter basis practically coincide. In such a basis the density matrices $\rho^{0}$ Eq.(\ref{e:rho}) and $\bar{\rho}^{0}$ Eq.(\ref{e:rhoanu}) as well as the corresponding mean-field hamiltonians  $h^{0}$ Eq.(\ref{e:mfnu}) and $\bar{h}^{0}$ Eq.(\ref{e:mfanu}) are diagonal. Therefore Eq.(\ref{e:hf}) is satisfied. This is our initial "stationary"  $(\rho^{0}, \bar{\rho}^{0})$ solution, until the Hamiltonian time-dependence drives the system out of equilibrium. At this point of the evolution  the density matrix  off-diagonal elements become nonzero. They  are the small amplitude deviations
$\rho'$ and $\bar{\rho}'$ ($ \rho'_{ij}$ and $\bar{\rho}'_{ij}$in components) of Eq.(\ref{e:rho1}). These satisfy Eqs.(\ref{e:tdanu}-\ref{e:RPA}) if the system has a small amplitude collective motion, or undergoes an instability. 
In this context one can assign to the quantities $\tilde{k}_k - \tilde{k}_l$  
the difference between the neutrino matter\footnote{Note that here we employ the terminology "matter" eigenvalues although they are obtained through the diagonalization of the Hamiltonian including mixings, neutrino interaction with matter and neutrino self-interaction.} eigenvalues at $t_0$ that are obtained by diagonalizing the mean-field Hamiltonian Eqs.(\ref{e:mfnu}) and (\ref{e:mfanu})
\begin{equation}\label{e:kvalues}
\tilde{h} =  U^{\dagger} h U = diag(\tilde{k}_i) ~~  \tilde{\bar{h}} =  U^{\dagger} \bar{h} U = diag(\tilde{\bar{k}}_i)
\end{equation}
 where the tilde here indicates that we are in the 'matter' basis.
 As for the derivatives of $h$ and $\bar{h}$, these can be obtained from the general expression for the mean field which is \cite{Volpe:2013uxl}
\begin{equation}\label{e:gamma}
\Gamma_{ij} = \sum_{kl} v_{(ik,jl)} \rho_{lk}.
\end{equation}
The explicit calculation of the matrix elements $v_{(ik,jl)} $ Eq.(\ref{e:matel}) associated with neutrino interaction with matter and neutrinos,
 gives the mean fields $\Gamma_{ij}$ shown  in Eqs.(\ref{e:mf}) and (\ref{e:mfanu})  (see \cite{Volpe:2013uxl} for details).  
One gets for a given set of single-particle quantum labels $ij$ the derivatives with respect to $kl$ with $k > l$ and $\bar{k}$ $\bar{l}$ with $\bar{k} > \bar{l}$ 
 \begin{equation}\label{e:der1}
 {\delta h_{ij} \over {\delta \rho'_{kl}}}= v^{\nu\nu}_{(il,jk)} ~~ {\delta h_{ij} \over {\delta \bar{\rho}'_{lk}}} = v^{\nu\bar{\nu}}_{(i\bar{k},j\bar{l})} 
\end{equation}
with $\bar{k}$ ($\bar{l}$) referring here to the single-particle states for an incoming (outoing) antineutrino and an explicit dependence of the matrix elements on
the interacting particles is introduced for clarity. 
Note also that there is no contribution from the derivatives 
\begin{equation}\label{e:der2}
{\delta h_{ij}  \over {\delta \rho'_{lk}}}={\delta h_{ij}  \over {\delta \bar{\rho}'_{kl}}} = 0 
 \end{equation}
 with respect to  $\delta \rho'_{lk}$  and to $ \delta \bar{\rho}'_{kl}$ ($k > l$).
Therefore, while the neutrino Hamiltonian receives contributions from the vacuum, the matter and the neutrino-neutrino terms, the mean-field derivatives with respect to variations of the density matrix receive no contribution from the vacuum and the matter terms that are density-independent. 
By implementing expressions (\ref{e:der1}-\ref{e:der2}) in Eq.(\ref{e:tdanu}) one finally finds 
the eigenvalue equations in the small amplitude approximation\footnote{ In the study of atomic nuclei,
this  eigenvalue equation is known as the Random-Phase-Approximation (RPA). The analogue of the neutrino single-particle energy differences, $\tilde{k}_k - \tilde{k}_l$, are the particle-hole excitation energies; while the two-body matrix elements are calculated using nuclear effective interactions. As is well established, the inclusion of this two-body residual interaction is necessary to reproduce the measured excitation energies of
the collective modes of atomic nuclei, known as the Giant Resonances.} 
\begin{align}\label{e:rpanu}
\omega \rho'_{ij}  &=  \sum_{k<l}
\{ (\tilde{k}_k - \tilde{k}_l)\delta_{ki} \delta_{jl} \rho'_{kl} \\
& \quad + 
(\rho^{0}_{j} - \rho^{0}_{i}) \left[v^{\nu\nu}_{(il,jk)} \rho'_{kl} + v^{\nu\bar{\nu}}_{(ik,jl)} \bar{\rho}'_{lk} \right] \}  \nonumber \\
\omega \bar{\rho}'_{ij}  &=  \sum_{k<l}
\{ (\tilde{\bar{k}}_l - \tilde{\bar{k}}_k)\delta_{kj} \delta_{il} \bar{\rho}'_{lk} \nonumber \\
& \quad + 
(\bar{\rho}^{0}_{j} - \bar{\rho}^{0}_{i}) \left[v^{\bar{\nu}\nu}_{(il,jk)} \rho'_{lk} + v^{\bar{\nu}\bar{\nu}}_{(il,jk)} \bar{\rho}'_{kl} \right] \}.Ê \nonumber
\end{align}
Again, equations equivalent to (\ref{e:rpanu}) hold for $\omega^*\rho'^{*}_{ij}$ and  $\omega^*  \bar{\rho}'^{*}_{ij} $.
Their solution determines the frequencies of the collective modes of  our system around the "stationary" solution. Note that such frequencies differ
from the single-particle energy differences, which are $\tilde{k}_k - \tilde{k}_l$ for neutrinos and $\tilde{\bar{k}}_k - \tilde{\bar{k}}_l$  for antineutrinos, by an amount that depends upon the two-body  interaction matrix elements.
In particular, the presence of complex frequencies indicates that the system has become unstable. 
The set of equations (\ref{e:tdanu}-\ref{e:RPA}) and (\ref{e:rpanu}) is our most general result of applying the linearization procedure to the $\nu$ mean-field equations in media.

We would like to emphasize the generality of the linearized equations we have derived. First, in our formulation we have not fixed the number of neutrinos, so that our equations can be used to study instabilities for an arbitrary number of families. This requires the use of the corresponding density matrix expression equivalent to Eqs.(\ref{e:rho}) and (\ref{e:rhoanu}) for a number of $N$ neutrinos instead of three. Second, Eqs.(\ref{e:lin1}) or (\ref{e:rpanu}) can be used to investigate the occurence as well as the precise location of flavor instabilities since all the contributions of the neutrino Hamiltonian have been retained in our derivation.
Third, while in our considerations $t_0$ has been taken to be the initial
time at the neutrinosphere, the procedure employed here can be used to linearize the neutrino equations of motion 
at any time of the evolution in a core-collapse supernova. 
To this aim, in Eqs.(\ref{e:linh}-\ref{e:tda}) one should take as the density matrix at the starting point  the one in the 'matter' basis.
Note that Ref.\cite{Galais:2011jh} has investigated the effects of the neutrino self-interaction in supernovae in such a basis. 
By definition, this is the basis that instantaneously diagonalizes the neutrino Hamiltonian, so that the stationary condition Eq.(\ref{e:hf}) is satisfied. In the small amplitude approximation, $\rho_{ij}'$ are the off-diagonal elements of the density matrix in such a basis. The eigenvalues 
$\tilde{k}_k - \tilde{k}_l$ are then the matter eigenvalues evaluated at the given $t_0$. Therefore, eigenvalue equations of the same type as Eq.(\ref{e:tdanu}) can be derived
to establish the presence of an instability at any time of the neutrino evolution. Finally, the problem of solving the linearized equations (\ref{e:tdanu})-(\ref{e:RPA}) can be replaced by the determination of the eigenvalues of the stability matrix $S$ (\ref{e:Smat}). While such a matrix remains positive definite for collective stable modes, its eigenvalues become complex in presence of instabilities.

\subsection{Eigenvalue equations in explicit form}
\noindent
The linearized equations are here written in a more explicit form for the case of neutrinos in a core-collapse supernova.
To this aim  
the expressions of the matrix elements $v_{(il,jk)}$ are needed corresponding to different scattering processes due to the (anti)neutrino charged- and neutral-current interaction 
with electrons, protons, neutrons and (anti)neutrinos of all flavors. The detailed derivation of the relevant matrix elements can be found in \cite{Volpe:2013uxl}. One can associate to the particle that is propagating for example the
 incoming and outgoing single-particle states $ij$ and to the particle of the background $kl$. Since only forward scattering is considered, $ij$ and $kl$ correspond to states with same momentum and helicity.  Under these assumptions one gets for Eq.(\ref{e:matel})
\begin{equation}\label{e:matelex} 
v_{(il,jk)} =  \left( 1- \hat{\vec{p}} \cdot \hat{\vec{k}}\right),
\end{equation}
with $ \hat{\vec{p}}$ and $\hat{\vec{k}}$ the momenta of the background and of the propagating particle, respectively.
For three flavors,  in the flavor basis Eqs.(\ref{e:rpanu}) give the following eigenvalue equations\footnote{Note that, to avoid confusing notations, from now we will denote the off-diagonal elements 
of the density matrices Eqs.(\ref{e:rho}) and (\ref{e:rhoanu}) as $\rho^{\nu_{\alpha}\nu_{\beta}}_{\vec{k}}$ instead of $\rho_{\nu_{\alpha}\nu_{\beta}, \vec{k}}$.} 
\begin{align}\label{e:lin2}
\omega \rho_{\vec{k}}^{\nu_{\alpha}\nu_{\beta}} & =  
(\tilde{k}_{\nu_{\alpha},\vec{k}} - \tilde{k}_{\nu_{\beta},\vec{k}}) \rho_{\vec{k}}^{\nu_{\alpha}  \nu_{\beta}} + \sqrt{2} G_F 
(\rho_{\vec{k}}^{\nu_{\beta}  \nu_{\beta}}- \rho_{\vec{k}}^{\nu_{\alpha}  \nu_{\alpha}})  \nonumber \\ 
& \quad  \times \sum_{\underline{\alpha}}  \int{{d^3 \vec{p}}\over{(2\pi)^3}}(1-\hat{\vec{k}} \cdot \hat{\vec{p}})
 (\rho_{\underline{\alpha},\vec{p}}^{\nu_{\alpha}\nu_{\beta}} -  \bar{\rho}_{\underline{\alpha},\vec{p}}^{*\nu_{\alpha}\nu_{\beta}}),
\end{align}
concerning the off-diagonal elements of the density matrix
Eq.(\ref{e:rho}), with $\alpha \neq \beta$ and $\alpha, \beta = e, \mu, \tau$. The quantity
 $ \rho_{\vec{k}}^{\nu_{\beta}  \nu_{\beta}}- \rho_{\vec{k}}^{\nu_{\alpha}  \nu_{\alpha}}$ is the difference of the diagonal matrix elements of the density matrix.
 Equation (\ref{e:lin2}) is associated with a given initial condition for $\rho$, $\rho_{\underline{\nu}_{\alpha}}$.
Analogous equations are valid for the $ \rho_{\vec{k}}^{*\nu_{\alpha}\nu_{\beta}}$ with eigenfrequencies $\omega^*$, as well as for the antineutrino counterparts $\bar{\rho}_{\vec{k}}^{\nu_{\alpha}\nu_{\beta}}$ and $\bar{\rho}_{\vec{k}}^{* \nu_{\alpha}\nu_{\beta}}$.  

In order to present our results in a more explicit form, we
consider as an example the "bulb model" for the neutrino emission at the neutrinosphere \cite{Duan:2010bg}. In this model one assumes both the spherical and the cylindrical symmetries (see Figure 1). The neutrinosphere corresponds to a sharp sphere of radius $R$.
\begin{figure}[!]
\includegraphics[scale=0.3]{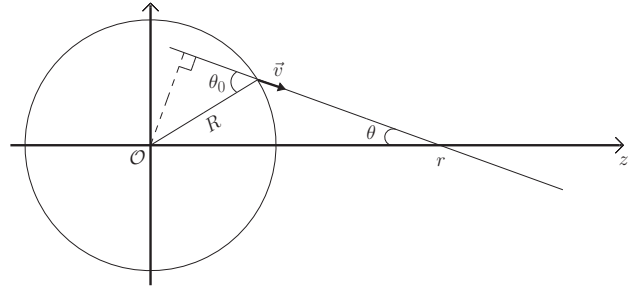}
\caption{Schematic picture of a neutrino propagating from the neutrinosphere, interacting with a neutrino from the background at a distance $r$ in a core-collapse supernova. The neutrinosphere is assumed to be a sharp sphere of radius $R$. The neutrino emission point is characterised by the variable $\theta$. The "bulb" model assumes both cylindrical and spherical symmetries. In such a model all neutrinos with $\alpha$ flavor, with same energy $k$ and $\theta$ undergo the same flavor history.  }
\label{fig:bulb}
\end{figure}
A neutrino or antineutrino flavor content at a given radius $r$ within a core-collapse supernova depends on its energy $k$, on $r$ and on the point of emission at the neutrinosphere. The latter can be characterized e.g. by the variable $ \theta  \in [0,\theta_{max}]$ with $\theta_{max}  = arcsin(R/r)$ (see \cite{Duan:2010bg}). All neutrinos born with a flavor $\nu_{\alpha}$ with same $k$ and $u$ have the same evolution  up to the distance $r$. 
Let us write Eqs.(\ref{e:lin2}) for the case of three-flavors within the "bulb" model. We introduce the velocity vector $\vec{v}$ with $|\vec{v}|=1$ to describe the motion of a (anti)neutrino with trajectory angle $ \theta$ with respect to the symmetry axis, $v = cos \theta$. The corresponding equation of motion Eq.(\ref{e:mf}) is
\beq\label{e:mf2}
i {d{\rho} \over {dr}}  = {1 \over {v}}  [h(\rho),\rho ] 
\eeq
with Eq.(\ref{e:mfnu}) now being
\begin{align}\label{e:mfnuex}
 h(\rho) &= UH_mU^{\dagger} +  H_{mat} + \sqrt{2} G_F \\
 & \times \sum_{\underline{\alpha}} \int {{p^2d p dv' }\over{(2 \pi)^2 }} (\rho_{\underline{\alpha},p,v'} - \bar{\rho}^*_{\underline{\alpha},p,v'}) 
 \left( 1 - v \cdot v'  \right) ,  \\ \nonumber
\end{align}
where $v'$ refers to the background particle; and similarly for Eqs.(\ref{e:mfa}-\ref{e:mfanu}).
With these assumptions the eigenvalue equations (\ref{e:lin2}) become at the single particle level\footnote{Note that  for the sake of clarity here we simplify the notation
and replace $\nu_{\beta'} \nu_{\beta}$ with $\beta' \beta$.}
\begin{align}\label{e:lin2fl}
\omega \rho_{\underline{\alpha}',\vec{k}}^{\beta' \beta} & =  (\tilde{k}_{\beta',\vec{k}} - \tilde{k}_{\beta,\vec{k}}) \rho_{\underline{\alpha}',\vec{k}}^{\beta' \beta} \nonumber \\ 
& + {\sqrt{2} G_F \over{2 \pi R^2}} (\rho_{\underline{\alpha}, \vec{k}}^{\beta' \beta'}  - \rho_{\underline{\alpha}, \vec{k}}^{\beta \beta})  \sum_{\underline{\alpha}}  
 \int d p dv' \nonumber \\ 
 & \quad  \times  ( {1 \over {v}} - v')  \Big[ \rho_{\underline{\alpha}, \vec{p}}^{\beta' \beta}   f_{FD}(p){L_{\underline{\alpha}}\over { \langle  E_{\underline{\alpha}} \rangle}}  -  \bar{\rho}_{\underline{\alpha},\vec{p}}^{* \beta' \beta} \bar{f}_{FD}(p)  {L_{\underline{\bar{\alpha}}}\over { \langle  E_{\underline{\bar{\alpha}}} \rangle}}  \Big]  , 
\end{align}
and similarly for antineutrinos. 
The quantities $L_{\underline{\alpha}}$ are the flux of neutrinos produced in the $\alpha$ flavor at the neutrinosphere and  $ \langle  E_{\underline{\alpha}} \rangle$
is the corresponding average neutrino energy. The functions $f_{FD} $, $\bar{f}_{FD}$ are Fermi-Dirac distributions for neutrinos and antineutrinos respectively,
depending on the two parameters ($T_{\nu},\eta_{\nu}$)
\begin{align}\label{e:FD}
f_{FD}(p) \equiv {1 \over {F_2(\eta_{\nu})T^3_{\nu}}} {p^2Ê\over {e^{(p/T_{\nu} -\eta_{\nu})}+1}}
\end{align}
where $T_{\nu}$ and $\eta_{\nu}$. The functions $F_k(\eta_{\nu})$ are 
\begin{align}\label{e:Fk}
F_k(\eta_{\nu}) \equiv \int^\infty_0  {x^k dx \over {e^{(x-\eta)} +1}}
\end{align}

By considering a discretized version of the integral in equations (\ref{e:lin2fl}), with respect to energy $p$ and to $v$,
the eigenvalue equations for neutrinos and antineutrinos can be cast together in the matrix form 
Eq.(\ref{e:lin1}) 
\begin{align}\label{e:rpanuv}
\omega 
\left(
\begin{array}{l}
\rho_{ \underline{\alpha}' }^{\beta' \beta}(\vec{k}) \\
 \\
\bar{\rho}_{ \underline{\alpha}' }^{*\beta' \beta}(\vec{k}) \\  
\end{array}
\right)
& =
 \left(
\begin{array}{ll}
A_{\underline{\alpha}' \underline{\alpha} }^{\beta' \beta}(\vec{k},\vec{p}) & B_{\underline{\alpha}' \underline{\alpha} }^{\beta' \beta}(\vec{k},\vec{p})\\
& \\
\bar{B}_{\underline{\alpha}' \underline{\alpha} }^{\beta' \beta}(\vec{k},\vec{p}) & \bar{A}_{\underline{\alpha}' \underline{\alpha} }^{\beta' \beta}(\vec{k},\vec{p})\\
\end{array} 
\right)
\left(
\begin{array}{l}
\rho_{ \underline{\alpha} }^{\beta' \beta}(\vec{p})  \\
\\
\bar{\rho}_{ \underline{\alpha} }^{*\beta' \beta}(\vec{p})  \\  
\end{array}
\right) 
  \\ \nonumber
\end{align}
where each element of the vectors is to be understood as a vector 
of dimension equal to $\tilde{n}=N \times n_p \times n_v$ with $N, n_p, n_v $ the number of neutrino families, neutrino energies and angles; while
each element of the matrix  is to be understood as a matrix of dimension $\tilde{n} \times \tilde{n}$. 
The corresponding components are
\begin{align}\label{e:S2f2}
A_{(\underline{\alpha}' \beta' \vec{k}; \underline{\alpha} \beta \vec{p} )} & =
 (\tilde{k}_{\beta',\vec{k}} - \tilde{k}_{\beta,\vec{k}}) \delta_{\underline{\alpha}'\underline{\alpha}} 
\delta_{\vec{k}\vec{p}} \delta_{\beta'\beta} + (\rho^{\alpha \alpha} - \rho^{\alpha' \alpha'}) \\ \nonumber
& \quad   \mu   D(v,v')  f_{FD}(p) {L_{\underline{\alpha}}\over { \langle  E_{\underline{\alpha}} \rangle} } \Delta p \Delta v' \\ \nonumber
B_{(\underline{\alpha}' \beta' \vec{k}; \underline{\alpha} \beta \vec{p} )} & = - (\rho^{\alpha \alpha} - \rho^{\alpha' \alpha'}) 
  \mu   D(v,v')  \bar{f}_{FD}(p) {L_{\underline{\bar{\alpha}}}\over { \langle  E_{\underline{\bar{\alpha}}} \rangle}} \Delta p \Delta v' \\ \nonumber
 \bar{B}_{(\underline{\alpha}' \beta' \vec{k}; \underline{\alpha} \beta \vec{p} )} & = - (\bar{\rho}^{\alpha \alpha} - \bar{\rho}^{\alpha' \alpha'}) 
 \mu    D(v,v') f_{FD}(p) {L_{\underline{\alpha}}\over { \langle  E_{\underline{\alpha}} \rangle} }\Delta p \Delta v'  \\ \nonumber
 \bar{A}_{(\underline{\alpha}' \beta' \vec{k}; \underline{\alpha} \beta \vec{p} )} & =
 (\tilde{\bar{k}}_{\beta',\vec{k}} - \tilde{\bar{k}}_{\beta,\vec{k}}) \delta_{\underline{\alpha}'\underline{\alpha}} 
\delta_{\vec{k}\vec{p}} \delta_{\beta'\beta} \\ \nonumber
& \quad + (\bar{\rho}^{\alpha \alpha} - \bar{\rho}^{\alpha' \alpha'}) \mu D(v,v') \bar{f}_{FD}(p) {L_{\underline{\bar{\alpha}}}\over { \langle  E_{\underline{\bar{\alpha}}} \rangle}}\Delta p \Delta v' \\ \nonumber
 \end{align}
 with $\Delta p \Delta v'$  being the energy and angular steps, while the coefficient $\mu$ is
 \beq\label{e:cons}
 \mu = { \sqrt{2} G_F \over{2 \pi R^2}}
 \eeq
 and the angular term is
  \beq \label{e:ang}
 D(v,v')= \left( {1 \over {v}} - v' \right) 
 \eeq
Equations (\ref{e:rpanuv}) and (\ref{e:S2f2})  identify the stability matrix (\ref{e:Smat}).

In Ref.\cite{Banerjee:2011fj} a different linearization procedure is applied to the neutrino mean-field equations in supernovae. The eigenvalue equation derived within the two-flavor framework is used to investigate the occurence of instabilities. Note that the result in \cite{Banerjee:2011fj} cannot be employed to identify the instability precise location since the eigenvalue equations are obtained, in particular, by neglecting the neutrino mixings and getting rid of the matter term by going to the comoving frame. In fact, it has been shown
 that instabilities are triggered because of the presence of mixings in the single-angle case \cite{Hannestad:2006nj,Galais:2011jh} as well as in multiangle calculations 
 \cite{Duan:2010bf}. If we make the same approximations\footnote{Note that the derivation in Ref.\cite{Banerjee:2011fj} also used a large $r$ limit of the Hamiltonian besides the assumptions already mentioned.} as those done Ref.\cite{Banerjee:2011fj}, our Eqs.(\ref{e:lin2}) for two flavors reduce to the eigenvalue equation (33) of Ref.\cite{Banerjee:2011fj}.

\section{Case with neutrino-antineutrino pairing correlations}
\noindent
One of our goals is to apply the procedure described in Section III.A to linearize the extended neutrino equations of motion,  that
include the neutrino-antineutrino correlations Eqs.(\ref{e:tddm}), around a "stationary" solution. Before doing that we focus on the extended Hamiltonian
and its diagonalization, to gain better insight on the kind of quantum states we are dealing with.

\subsection{Extended Hamiltonian}
\noindent
The generalized Hamiltonian ${\cal H} $ Eq.(\ref{e:hex}) is derived in \cite{Volpe:2013uxl} by retaining linked contributions of the $\nu$-$\bar{\nu}$ type, in the evolution equation of the two-body correlation function.
An alternative procedure to obtain ${\cal H} $ is given here, for the particular case when one considers that the expectation values are performed on the ground state  $| \phi \rangle $ of the system including $\nu$-$\bar{\nu}$ pairing correlations. Let us start from  the Hamiltonian Eq.(\ref{e:Hf}), written in second quantization, describing our system of neutrinos and antineutrinos interacting with a matter, a $\nu$ and a $\bar{\nu}$ background:
\begin{align}\label{e:Hsq}
H^f & = \sum_{k,k'} \left[ \epsilon_{k,k'} a^{\dagger}_k a_{k'} +  \bar{\epsilon}_{k,k'} b^{\dagger}_k b_{k'} \right] + {1 \over{2}}  \sum_{mnij} v_{mn,ij} a^{\dagger}_{m} a^{\dagger}_{n} a_{j} a_{i}  \nonumber \\
& + {1 \over 2}  \sum_{mnij} v_{mn,ij} b^{\dagger}_{m} b^{\dagger}_{n} b_{j} b_{i} + {1 \over{2}}  \sum_{mnij} v_{mn,ij} a^{\dagger}_{m} b^{\dagger}_{n} b_{j} a_{i}  
\end{align} 
where the first two-body term corresponds to neutrino-electron and neutrino-neutrino scattering, the second to antineutrino-antineutrino scattering 
and the third interaction term corresponds to anti-neutrino electron and $\nu-\bar{\nu}$ scattering.   Therefore the interaction terms involving $\nu$ or $\bar{\nu}$ run over all flavors.

By applying the Wick theorem, the  Hamiltonian can be developed  with respect to the ground state  $| \phi \rangle $ of the system including $\nu$-$\bar{\nu}$ pairing correlations to obtain a mean field approximation. For the one-body term in (\ref{e:Hsq}) one obtains\footnote{Note that the expectation values $\langle \phi |  a^{\dagger}_k a_{k'}  |  \phi \rangle$ and $\langle \phi | b^{\dagger}_k b_{k'}  |  \phi \rangle $ are not zero. This is because
$| \phi \rangle$ is not the vacuum with respect to the particle operators $a$ and $b$. If  $| \phi \rangle$ is the corresponding vacuum then $a^{\dagger}_k a_{k'} = : a^{\dagger}_k a_{k'} :$ and $b^{\dagger}_k b_{k'} = : b^{\dagger}_k b_{k'} : $ hold.}
\begin{align}\label{e:ob}
a^{\dagger}_k a_{k'} & = \langle \phi |  a^{\dagger}_k a_{k'}  |  \phi \rangle + : a^{\dagger}_k a_{k'} :  = \rho_{kk'}  +    : a^{\dagger}_k a_{k'} :    \nonumber \\
b^{\dagger}_k b_{k'} & = \langle \phi | b^{\dagger}_k b_{k'}  |  \phi \rangle + : b^{\dagger}_k b_{k'} :  = \bar{ \rho}_{kk'}  + : b^{\dagger}_k b_{k'} : 
\end{align}
with $: :$ referring here to normal ordering;  while the two-body interaction terms give:
\begin{align}\label{e:tb}
b^{\dagger}_{m} b^{\dagger}_{n} b_{j} b_{i} & = \bar{ \rho}_{im} \bar{ \rho}_{jn} - \bar{ \rho}_{jm} \bar{ \rho}_{in} 
+ \bar{ \rho}_{jn} : b^{\dagger}_m b_{i} : \nonumber \\
& + \bar{ \rho}_{im} : b^{\dagger}_n b_{j} :  - \bar{ \rho}_{jm} : b^{\dagger}_n b_{i} : - \bar{ \rho}_{in} : b^{\dagger}_m b_{j} :  \nonumber \\
& +  : b^{\dagger}_m  b^{\dagger}_n b_{j} b_i : 
\end{align}
and 
\begin{align}\label{e:tb2}
a^{\dagger}_{m} b^{\dagger}_{n} b_{j} a_{i} & =  \rho_{im} \bar{ \rho}_{jn} + \kappa^*_{nm} \kappa_{ij} + \rho_{im} : b^{\dagger}_n b_{j} :  \nonumber \\
&  +   \bar{ \rho}_{jn} : a^{\dagger}_m a_{i} : 
+ \kappa_{nm}^* : b^{\dagger}_j a_{i} : + \kappa_{ij} : a^{\dagger}_m b_{n} :
\end{align}
A similar expression as  Eq.(\ref{e:tb}) can be written for the two-body interaction term $a^{\dagger}_{m} a^{\dagger}_{n} a_{j} a_{i}$ in Eq.(\ref{e:Hsq}).
Note that in Eqs.(\ref{e:tb}-\ref{e:tb2}) the terms of the type $\langle \phi |  a_k a_{k'}  |  \phi \rangle$ and $\langle \phi |  b_k b_{k'}  |  \phi \rangle$ (and their hermitian conjugates) have not been retained since such contributions are not included in our extended Hamiltonian (see Eq.(\ref{e:ex}) and Ref.\cite{Volpe:2013uxl}) 
Putting Eqs.(\ref{e:ob}-\ref{e:tb2}) together and neglecting the contributions from the normal order products of four operators
one gets the mean-field contribution from the Hamiltonian (\ref{e:Hsq}) :
\begin{align}\label{e:Hmf}
H^f = H_{gs} + H_{MF}
\end{align}
The first term  $H_{gs}= \langle\phi | H^f | \phi \rangle $ corresponds to the ground state energy :
\beq\label{e:Hgs}
H_{gs} = tr(\epsilon  \rho +  \bar{\epsilon} \bar{\rho}) + {1 \over 2} tr(\Gamma^{\nu} \rho + \Gamma^{\bar{\nu}} \bar{\rho}+ \Gamma^{\nu\bar{\nu}} \bar{\rho}) + {1 \over 2} tr(\Delta  \kappa^*)
\eeq
where $tr$ indicates that we are tracing over single-particle states , $\Gamma^{\nu}$ is the mean field Eq.(\ref{e:gamma}) acting on neutrinos
$\Gamma^{\nu} = \Gamma^{\nu e} + \Gamma^{\nu\nu}$ that receives contributions from the electron and neutrino backgrounds, 
$\Gamma^{\bar{\nu}} = \Gamma^{\bar{\nu}e} + \Gamma^{\bar{\nu}\bar{\nu}} $ is the one acting on anti-neutrinos due to the electron and the antineutrino backgrounds, while 
$\Gamma^{\nu\bar{\nu}} $ is the mean field due to antineutrinos and acting on neutrinos. 
The second term in (\ref{e:Hmf}) is the mean field $H_{MF}= \sum_{kk'} \left[H_{MF} \right]_{kk'}$ where each term can be cast a matrix form:
\beq\label{e:Hdev}
\begin{array}{lccc}
\left[H_{MF} \right]_{kk'} & = & : ( a^{\dagger} & b )_k \\  
\end{array} 
 \Big(
\begin{array}{lc}
h'  &   \Delta \\
 \Delta^{\dagger} &  - \bar{h'}^* \\  
\end{array} \Big)_{kk'}
\Big(
\begin{array}{l}
a  \\
b^{\dagger} \\  
\end{array} \Big)_{k'} :
\eeq
The neutrino and antineutrino mean fields $h=\epsilon +   \Gamma^{\nu\nu} + (\Gamma^{\nu e} + \Gamma^{\nu\bar{\nu}})/2 $\footnote{Note that the contributions of the type $\Gamma^{\nu\nu}$ and $\Gamma^{\bar{\nu}\bar{\nu}} $ are divided by a factor of 2 if the neutrinos or antineutrinos do not refer to equal particles.}  and 
$\bar{h}=\bar{\epsilon}  + \Gamma^{\bar{\nu}\bar{\nu}} +  (\Gamma^{\bar{\nu}e} +  \Gamma^{\bar{\nu}\nu})/2 $, are now defined as $h'=h-\lambda$  and $\bar{h'}=\bar{h}^*-\bar{\lambda}$ with $\lambda$ and $\bar{\lambda}$ Lagrange multipliers. 
These  are added to implement the condition that  the expectation value of the number operator over the state $| \phi \rangle $ is conserved on average. 
From Eq.(\ref{e:Hdev}) one can see that this procedure leads to an extended Hamiltonian consistent with $\cal{H}$  Eq.(\ref{e:hex}).

\subsection{Generalized Bogoliubov-Valatin transformation}
\noindent
The extended neutrino Hamiltonian (\ref{e:Hdev}) can be put in a diagonal form
by performing a generalized Bogoliubov 
transformation from the particle to the quasi-particle operators of the following kind :
\beq\label{e:genBog1m}
\Big(
\begin{array}{l}
 \alpha_{k}  \\
 \alpha_{\bar{k}}^{\dagger} \\  
\end{array} \Big)
 = {\cal X}^{\dagger}
\Big(
\begin{array}{l}
 a_{k}  \\
 b_{\bar{k}}^{\dagger} \\  
\end{array} \Big)
\eeq 
with the unitary matrix ${\cal X}$ defined as:
\beq\label{e:W}
{\cal X}^{\dagger} = \Big(
\begin{array}{lcl}
 Z^{\dagger} & W^{\dagger} \\
  V^T& U^T 
\end{array} \Big)
\eeq 

\noindent
In component form, the transformation (\ref{e:genBog1m}) reads :
\beq\label{e:genBog1}
\left\{
\begin{array}{lcl}
\alpha_{k}^{\dagger} & = & \sum_l Z_{lk} a^{\dagger}_l  + W_{lk} b_{\bar{l}} \\
\alpha_{\bar{k}} & = & \sum_l V^*_{lk} a^{\dagger}_l  + U^*_{lk} b_{\bar{l}} 
\end{array} \right.
\eeq 
and 
\beq\label{e:genBog2}
\left\{
\begin{array}{lcl}
\alpha_{k'}& = & \sum_{l'} Z^*_{l'k'} a_{l'} + W^*_{l'k'} b^{\dagger}_{\bar{l}'}  \\
\alpha^{\dagger}_{\bar{k}'} & = & \sum_{l'} V_{l'k'} a_{l'} + U_{l'k'} b^{\dagger}_{\bar{l}'}
\end{array} \right.
\eeq 
The generalized Bogoliubov-Valatin transformation comprises two sets of quasi-particle operators, 
namely $\alpha_{k}, \alpha_{k}^{\dagger}$ and $\alpha_{\bar{k}}, \alpha_{\bar{k}}^{\dagger}$, 
with the indices running over the $k$-particle  and $\bar{k}$-antiparticle quantum states, respectively. One sees that
Eqs.(\ref{e:genBog1}-\ref{e:genBog2}) are linear combinations of the particle and of the antiparticle operators, thus combining a neutrino single-particle state with its $CP$-conjugate.  Ref.\cite{Pehlivan:2011hp} has shown that the Hamiltonian describing a system of self-interacting neutrinos and antineutrinos is equivalent to the reduced BCS Hamitonian. The authors have introduced Bogoliubov transformations for neutrinos (or antineutrinos)
combining operators associated with different flavor (or mass) states. 

The $U, V, W, Z$ matrices  in Eqs.(\ref{e:genBog1m}) and (\ref{e:genBog1}-\ref{e:genBog2}) are constrained by requiring that the quasi-particle operators satisfy the same anti-commutation rules than the particle and antiparticle operators. Thus they have to satisfy the following relations:
\beq\label{e:unitcond}
\begin{array}{lcl}
Z Z^{\dagger} + V^*V^T & = &  \bf{1}    \\
WW^{\dagger} + U^*U^T & = &    \bf{1}  \\
Z^{\dagger}Z + W^{\dagger}W & = &  \bf{1}    \\
V^{\dagger}V + U^{\dagger}U & = &  \bf{1}     \\
ZW^{\dagger} + V^{*}U^T & = & 0    \\
V^{T}Z + U^{T}W & = & 0   
\end{array} 
\eeq 
that are ensured by the unitarity of the ${\cal X}^{\dagger}$ transformation Eq.(\ref{e:genBog1m}). 

A special case for the transformations (\ref{e:genBog1}-\ref{e:genBog2}) is represented by the special Bogoliubov-Valatin transformations:
\beq\label{e:spBog1}
\left\{
\begin{array}{lcl}
\alpha_{\bar{k}} & = &  v^*_{k} a^{\dagger}_k + u^*_{k} b_{\bar{k}}  \\
\alpha_{k}^{\dagger} & = & z_{k} a^{\dagger}_k +  w_{k} b_{\bar{k}} 
\end{array} \right.
\eeq 
and the corresponding relations for the hermitian conjugate operators.
Relations (\ref{e:unitcond}) imply that the $u_{k}$, $v_{k}$ , $w_{k}$, $z_{k}$ coefficients have to satisfy the following relations: $|u|^2_{k} + |v|^2_{k} =1$, $|w|^{'2}_{k} + |z|^{'2}_{k} =1$, $|u|^2_{k} + |w|^2_{k} =1$, $|z|^2_{k} + |v|^2_{k} =1$ and $v_{k}z_{k} + w_{k}u_{k}=0$, $z_{k}w^*_{k}+v_{k}^*u_{k}=0$.
In the limit of vanishing abnormal mean field, since $u_{k} \rightarrow 1$, $v_{k} \rightarrow 0$ and $w_{k} \rightarrow 1$, $z_{k} \rightarrow 0$, the quasi-particle operators $\alpha^{\dagger}_{\bar{k}}$ and $\alpha_{\bar{k}}$ tend to the antineutrino operators $b^{\dagger}_{\bar{k}}$ and $b_{\bar{k}}$; while the quasi-particle operators $\alpha_{k}^{\dagger}$ and $\alpha_{k}$ tend to the neutrino operators $a^{\dagger}_k$ and $a^{\dagger}_{k}$. In this case, the extended Hamiltonian reduces to the usual mean field.

\subsection{Quasi-particle eigenenergies}
\noindent
Using Eqs.(\ref{e:genBog1}-\ref{e:genBog2}), it is straightforward to show that  $H_{MF,qp} = {\cal X}^{\dagger}  H_{MF}  {\cal X} $, so that Eq.(\ref{e:Hdev}) becomes
\beq\label{e:Hdiag}
H^f_{qp} = H_{gs} + \sum_{k,\bar{k}} (\epsilon_{k} \alpha^{\dagger}_k \alpha_{k} + \bar{\epsilon}_{\bar{k}} \alpha^{\dagger}_{\bar{k}} \alpha_{\bar{k}}) 
\eeq 
with $\epsilon_{k}, \bar{\epsilon}_{k}$ the quasi-particle eigenenergies which can be determined through 
\beq\label{e:diag}
 \Big(
\begin{array}{lcl}
 Z^{\dagger} & W^{\dagger} \\
  V^T& U^T 
\end{array} \Big)
 \Big(
\begin{array}{lc}
h'  &   \Delta \\
 \Delta^{\dagger} &   -\bar{h'}^* \\  
\end{array} \Big)
 \Big(
\begin{array}{lcl}
 Z & V^* \\
  W & U^* 
\end{array} \Big) =
\Big(
\begin{array}{lcl}
 \epsilon & 0 \\
  0 & -\bar{\epsilon}^*
\end{array} \Big)
\eeq
giving the following eigenvalue equations:
\beq\label{e:hfb1}
 \Big(
\begin{array}{lc}
h'  &   \Delta \\
 \Delta^{\dagger} &   -\bar{h'}^* \\  
\end{array} \Big)
\Big(
\begin{array}{l}
Z  \\
W \\  
\end{array} \Big) =
 \Big(
\begin{array}{l}
Z  \\
W \\  
\end{array} \Big) 
\epsilon
\eeq
and 
\beq\label{e:hfb2}
 \Big(
\begin{array}{lc}
h'  &   \Delta \\
 \Delta^{\dagger} &  - \bar{h'}^* \\  
\end{array} \Big)
\Big(
\begin{array}{l}
V^*  \\
U^* \\  
\end{array} \Big) = -
\Big(
\begin{array}{l}
V^*  \\
U^* \\  
\end{array} \Big) 
\bar{\epsilon}^* 
\eeq
The solution of such equations allows to determine the quasi-particle energy eigenvalues $\epsilon$ and $\bar{\epsilon}$, as well as the $U, V, Z$ and $W$ of the transformations Eqs.(\ref{e:genBog1}-\ref{e:genBog2}). One can then express the normal and abnormal densities as a function of the quasi-particle operators,
according to the following relations:
\beq\label{e:Rqp}
\begin{array}{lcl}
\rho_{ll'}=\langle a^{\dagger}_{l'}a_l  \rangle  &=& \sum_{kk'} V_{l'k'} V_{lk}^* (1-\bar{f}_{kk'}) + Z_{l'k'}^* Z_{lk}  f_{kk'} \\ 
\bar{\rho}_{ll'}=\langle b^{\dagger}_{l'}b_l  \rangle  &=& \sum_{kk'} U_{l'k'} U_{lk}^* \bar{f}_{kk'} +  W_{lk}^* W_{l'k'} (1- f_{kk'}) \\  
\kappa_{ll'}=\langle b_{l'}a_l  \rangle  &=& \sum_{kk'} V_{l'k'}^* U_{lk} (1-\bar{f}_{k'k}) + W_{l'k'}^* Z_{lk}  f_{kk'}  
\end{array} 
\eeq
and similarly for $\kappa_{ll'}^*$, with $\bar{f}_{kk'}= \langle \alpha^{\dagger}_{\bar{k'}} \alpha_{\bar{k}}  \rangle  $ and $ f_{kk'}= \langle \alpha^{\dagger}_{k'} \alpha_{k}  \rangle $.
The stationary independent quasi-particle state that corresponds to the ground state of the Hamiltonian (\ref{e:Hdiag}) is the vacuum of the quasi-particle operators 
defined by $\alpha_k | \Phi \rangle =0$ and $\alpha_{\bar{k}}  |  \Phi \rangle =0$. In this case relations (\ref{e:Rqp}) reduce to :
\beq\label{e:Rqvac}
\begin{array}{lcl}
\rho_{ll'}  &=& (V^* V^T)_{ll'}  \\ 
\bar{\rho}_{ll'} &=& (U^* V^T)_{ll'}  \\  
\kappa_{ll'} &= &(V^* U^T)_{ll'}  
\end{array} 
\eeq

\section{The linearization of the extended evolution equations}
\noindent
We now apply our procedure to linearize the extended neutrino evolution equations (\ref{e:ex}-\ref{e:hex}) assuming that at initial time the system is approximately described by a "stationary" solution which satisfies
\beq\label{e:hfb}
[{\cal H}^{0},{\cal R}^{0}] = 0,
\eeq
with 
\beq\label{e:hfz}
{\cal H}^{0} = \left(
\begin{array}{cc}   
h^{0} & \Delta^{0} \\
\Delta^{0 \dagger} & -\bar{h}^{0*} \end{array}
\right)
\end{equation}
and
\begin{equation}\label{e:rez}
{\cal R}^{0} =
 \left(
\begin{array}{cc}   
 \rho^{0}  &  \kappa^{0}   \\
\kappa^{0 \dagger} & 1 - \bar{\rho}^{0*} 
\end{array}
\right).
\end{equation}
Eq.(\ref{e:hfb}) is equivalent to Eq.(\ref{e:tdhf}) in presence of $\nu\bar{\nu}$ pairing.
Following Eq.(\ref{e:rho1}), we then consider small amplitude variations around such a solution, that are given by
\beq\label{e:smallR}
\delta {\cal R} = {\cal R}^{0} + \delta {\cal R}'
\eeq
The quantity 
\beq\label{e:deltaR}
\delta {\cal R}' =
 \left(
\begin{array}{cc}   
 \delta \rho  & \delta \kappa   \\
\ \delta \kappa^{\dagger} &  - \delta \bar{\rho}^{*} 
\end{array}
\right).
\end{equation}
 involves the small amplitude variations of both the normal and the abnormal density 
\beq\label{e:smallk2}
\begin{array}{lcl} 
 \delta \rho(t) & = & \rho' e^{-i\omega t} +  \rho'^{\dagger}e^{i\omega^* t}  \\
 \delta \bar{\rho}(t) & = & \bar{\rho}' e^{-i\omega t} +  \bar{\rho}'^{\dagger}e^{i\omega^* t}  \\
\delta \kappa (t) & = & \kappa_+e^{-i\omega t} +  \kappa_-e^{i\omega t}. \\
\delta \kappa^* (t) & = & \kappa_-^*e^{-i\omega^* t} +  \kappa_+^*e^{i\omega^* t}.  
 \end{array}
\eeq
where independent amplitudes $\kappa_+ $ and $\kappa_-$ for positive and negative frequencies have been considered. 
Note the abnormal density Eq.(\ref{e:kmat}) is not hermitian.
One can develop the extended Hamiltonian ${\cal H}$ (\ref{e:hex}) around ${\cal H}^{0}$ 
\beq\label{e:linH}
{\cal H}({\cal R}) = {\cal H}^{0} + {{\delta {\cal H}}\over{\delta {\cal R}}}\Big|_{{\cal R}^{0}}{\delta {\cal R}} + \ldots
\eeq
which implies considering the derivatives of $h$ and $\bar{h}$ with respect to $\delta \bar{\rho}$  and  $\delta {\rho}$ Eqs.(\ref{e:linh2}),
as in Eq.(\ref{e:linh}), as well as the variations of the abnormal fields with respect to the abnormal densities
\begin{align}\label{e:de}
\delta \Delta = {\delta \Delta \over {\delta \kappa}}\Big|_{\kappa^0}\delta \kappa
\end{align}
\begin{align}\label{e:des}
\delta \Delta^* = {\delta \Delta^* \over {\delta \kappa^*}}\Big|_{\kappa^{0*}}\delta \kappa^*.
\end{align}
By implementing  $\delta {\cal R}$  Eq.(\ref{e:smallR})
and Eq.(\ref{e:linH}) on the {\it l.h.s.} and {\it r.h.s.} of Eq.(\ref{e:ex}), one gets 
\begin{align}\label{e:linH1}
i\delta \dot{\cal R}'& = [{\cal H}^{0}, {\cal R}^{0}] + [{{\delta {\cal H}}\over{\delta {\cal R}}}\Big|_{{\cal R}^{0}}{\delta {\cal R}}, {\cal R}^{0}] \\ \nonumber
& \quad + [{\cal H}^{0},\delta {\cal R}' ] + 
[ {{\delta {\cal H}}\over{\delta {\cal R}}}\Big|_{{\cal R}^{0}}{\delta {\cal R}},\delta {\cal R}' ] \\ 
\end{align}
By virtue of Eq.(\ref{e:hfb}) and retaining only the lowest order terms one obtains
\begin{align}\label{e:linH1}
i\delta \dot{\cal R}' & =  [{{\delta {\cal H}}\over{\delta {\cal R}}}\Big|_{{\cal R}^{0}}{\delta {\cal R}}, {\cal R}^{0}]  + [{\cal H}^{0},\delta {\cal R}'] 
\end{align}

Implementing ${\cal H}^{0}$ Eq.(\ref{e:hfz}), ${\cal R}^{0}$ Eq.(\ref{e:rez}), Eqs.(\ref{e:dh}-\ref{e:dhb}) and (\ref{e:de}-\ref{e:des}) for the derivatives of the normal and pairing mean fields, the eigenvalue equations for the component form Eqs.(\ref{e:tddm}) are obtained for the extended neutrino evolution equations. These read
\begin{align}\label{e:qrpa1}
\omega \rho'_{1,ij}  & = \sum_m \sum_{pq} \{ (h^0_{1,im} \delta_{p_1m} \delta_{q_1j} - h^0_{1,mj} \delta_{p_1i} \delta_{q_1m} \\ \nonumber
& \quad + v_{(iq_1,mp_1)} \rho^0_{1,mj} - v_{(mq_1,jp_1)} \rho^0_{1,im})\rho'_{1,p_1q_1} \\ \nonumber
& \quad  + (v_{(iq_2,mp_2)} \rho^0_{1,mj} - v_{(mq_2,jp_2)} \rho^0_{1,im} ) \rho'_{2, p_2 q_2}  \\ \nonumber
& \quad  + (\Delta^0_{im}\delta_{p_1j} \delta_{q_2m} - \kappa^0_{im}v^*_{(jm,p_1q_2)}) \kappa^*_{-,p_1,q_2} \\ \nonumber
& \quad + (v_{(im,p_1q_2)}\kappa^{0*}_{jm} - \Delta^{0*}_{jm} \delta_{p_1i} \delta_{q_2m}) \kappa_{+,p_1q_2} \} \\ \nonumber
\end{align}

\begin{align}\label{e:qrpa2}
\omega \rho'_{2,kl}  & = \sum_m \sum_{pq} \{ (h^0_{2,km} \delta_{p_2m} \delta_{q_2l} - h^0_{2,ml} \delta_{p_2k} \delta_{q_2m} \\ \nonumber
& \quad + v_{(kq_2,mp_2)} \rho^0_{2,ml} - v_{(mq_2,lp_2)} \rho^0_{2,km})\rho'_{2,p_2q_2} \\ \nonumber
& \quad  + (v_{(kq_1,mp_1)} \rho^0_{2,ml} - v_{(mq_1,lp_1)} \rho^0_{2,km} ) \rho'_{1, p_1 q_1}  \\ \nonumber
& \quad  + (\Delta^0_{mk}\delta_{p_1m} \delta_{q_2l} - \kappa^0_{mk}v^*_{(ml,p_1q_2)}) \kappa^*_{-,p_1,q_2} \\ \nonumber
& \quad + (v_{(mk,p_1q_2)}\kappa^{0*}_{ml} - \Delta^{0*}_{ml} \delta_{p_1m} \delta_{q_2k}) \kappa_{+,p_1q_2} \} \\ \nonumber
\end{align}

\begin{align}\label{e:qrpa3}
\omega \kappa_{+,ik}  & = \sum_m \sum_{pq} \{ (h^0_{2,1m} \delta_{p_1m} \delta_{q_2k} + h^0_{2,km} \delta_{p_1i} \delta_{q_2m} \\ \nonumber
& \quad - v_{(mk,p_1q_2)} \rho^0_{1,im} - v_{(im,lp_1q_2)} \rho^0_{2,km})\kappa_{+,p_1q_2} \\ \nonumber
& \quad  + (v_{(iq_1,mp_1)} \kappa^0_{mk} + v_{(kq_1,mp_1)} \kappa^0_{im}  \\ \nonumber
& \quad - \Delta^0_{mk} \delta_{p_1i} \delta_{q_1 m} ) \rho'_{1, p_1 q_1} + (v_{(iq_2,mp_2)}\kappa^{0*}_{mk}  \\ \nonumber
& \quad + v_{(kq_2,mp_2)} \kappa^0_{im}  - \Delta^0_{im} \delta_{p_2k} \delta_{q_2 m} ) \rho'_{2, p_2 q_2} \} \\ \nonumber
\end{align}

\begin{align}\label{e:qrpa3}
\omega \kappa_{+,ik}  & = \sum_m \sum_{pq} \{ (h^0_{2,1m} \delta_{p_1m} \delta_{q_2k} + h^0_{2,km} \delta_{p_1i} \delta_{q_2m} \\ \nonumber
& \quad - v_{(mk,p_1q_2)} \rho^0_{1,im} - v_{(im,lp_1q_2)} \rho^0_{2,km})\kappa_{+,p_1q_2} \\ \nonumber
& \quad  + (v_{(iq_1,mp_1)} \kappa^0_{mk} + v_{(kq_1,mp_1)} \kappa^0_{im}  \\ \nonumber
& \quad - \Delta^0_{mk} \delta_{p_1i} \delta_{q_1 m} ) \rho'_{1, p_1 q_1} + (v_{(iq_2,mp_2)}\kappa^{0*}_{mk}  \\ \nonumber
& \quad + v_{(kq_2,mp_2)} \kappa^0_{im}  - \Delta^0_{im} \delta_{p_2k} \delta_{q_2 m} ) \rho'_{2, p_2 q_2} \} \\ \nonumber
\end{align}
\noindent
where the dependence of the labels on particle 1 and particle 2 has been made explicit and indicates here neutrinos (for particle 1) and antineutrinos (for particle 2).
In general, in the particle  basis employed here, both the static and the variations of densities and mean fields
can have  nonzero diagonal and off-diagonal elements. If one considers the generalized Bogoliubov-Valatin transformation Eq.(\ref{e:genBog1m})
from the particle to the quasi-particle basis, both the generalized Hamiltonian and density are diagonal in such a basis as showed in section V. The condition given by Eq.(\ref{e:hfb}) is therefore automatically satisfied, if the system finds itself in a quasi-particle eigenstate at initial time. One can perform a linearization of the extended neutrino equations of motion in the quasiparticle basis, in the same way as
in the particle basis. However, in this case,  the normal and abnormal density variations Eqs.(\ref{e:smallk2}) will only involve nonzero off-diagonal components, as it has been the case in Section III. 

\section{Discussion and conclusions}
\noindent
In the present work we have investigated formal aspects of the extended mean field equations describing the evolution of neutrinos and antineutrinos traversing a medium.  The equations are extended because they
 include neutrino-antineutrino pairing correlations and a pairing mean field besides the contributions due to neutrino interactions with matter and neutrinos that are usually included. 
Two aspects have been investigated : the static solution of the extended Hamiltonian describing our neutrino system and the linearized version of the  equations of motion around such a static solution.

The study of the static solution of the extended Hamiltonian  has shown that our system of interacting neutrinos with
$\nu\bar{\nu}$ pairing correlations can be described as a system of independent quasi-particles.
The latter have the specificity of being a mixture of neutrinos and antineutrinos, with the constrain that the particle and antiparticle have opposite momentum. Such a  requirement on the momenta comes from assuming the system to be homogeneous, as pointed out in our previous work \cite{Volpe:2013uxl} where the extended equations are derived. Bogoliubov-Valatin transformations typically involve a combination a particle operator and its 
 $T$-conjugate operator (see e.g. \cite{Ring}), where $T$- stands for time reversal. In the language of many-body microscopic approaches, 
such a transformation mixes a particle with a hole state. In the present work the generalized Bogoliubov-Valatin transformation necessary to diagonalize our extended Hamiltonian 
are of different kind and consists of a combination of a particle operator with the one corresponding to its $CP$-conjugate particle.
In some respects,  the system we have been investigating  has similarities with the pairing between protons and neutrons in atomic nuclei (see e.g. \cite{Camiz:1966ra}). We have demonstrated that with generalized Bogoliubov-Valatin transformation our Hamiltonian becomes diagonal and 
have given the equations to determine the energy eigenvalues of the quasi-particle eigenstates. In the many-body language these are usually referred to as the Hartree-Fock-Bogoliubov equations. 

We have obtained a linearized version of the neutrino evolution equations with a procedure known in the study of many-body systems.
In our case, neutrinos and antineutrino density matrix variations around the  static solution play the role of the RPA forward and backward amplitudes. 
The eigenvalue equations and the corresponding stability matrix  involve 
neutrino and antineutrino interaction matrix elements, instead of the particle-hole particle-hole ones usually present in the RPA approach. In the study of atomic nuclei the solution of  small amplitude eigenvalue equations determines for example the excitation energies of giant resonances. Therefore one typically requires the stability matrix to be positive definite. On the other hand the occurrence of instabilities, when correlations become too large, is well known in the study of neutrino-less double beta-decay \cite{Elliott:2004hr}. Their appearance signals the breakdown of the small amplitude approximation inherent to the method. 
We have stressed the generality of the linearized equations derived in the mean-field case that can be used for an arbitrary number of neutrino families. We emphasize 
that our equations retain all contributions of the neutrino Hamiltonian which is an important aspect if one wants to identify the precise location of instabilities in flavor space. This is for example relevant if one wants to apply the linearized equations in nucleosynthesis studies, where it has been shown that the exact
location of flavor changes can determine completely different nucleosynthetic outcomes \cite{Duan:2010af}. 

A linearized version of the neutrino evolution equations with $\nu\bar{\nu}$ correlations has been derived. While the results we have furnished are in the particle basis, one could also derive eigenvalue equations in the small amplitude approximation directly in the quasi-particle basis. This would have the advantage that only the off-diagonal matrix elements for the normal and abnormal densities would have nonzero contributions. As we have been discussing, in some cases, an alternative way of searching for small amplitude stable or unstable modes, is through the stability matrix that can be derived by developing the energy density around a stationary solution. In particular, if  the generalized density matrix ${\cal R}$ is idempotent, density matrix variations around the ground state will keep being quasi-particle states. Therefore the stability matrix can be constructed directly from such variations \cite{Ring}. Note that in the neutrino case we have been considering here, the abnormal density considered is not skew symmetric and the  ${\cal R}$ is in general not idempotent, while in some cases $\kappa$ and $\rho$ might satisfy the necessary relations\footnote{These are $\rho^2 - \rho = - \kappa \kappa^{\dagger}$ and $\rho\kappa = \kappa \rho^*$.} to have $ {\cal R}^2= {\cal R}$. 

The methods and the equations presented in this manuscript can be of practical use  in realistic studies of neutrino flavor conversion in media, in particular core-collapse supernovae or in accretion-disks black hole scenarios. They can help identifying collective neutrino modes as well as the presence of spurious modes that come in as an artefact of the approximations made, as in the study of atomic nuclei, because of the breaking of symmetries. 
The ensemble of the results presented in this manuscript further uncover connections between formal aspects of the neutrino propagation in astrophysical environments and of many-body systems like atomic nuclei, clusters and condensed matter. 

\begin{acknowledgments}
\noindent
C. V. would like to thank J. Serreau for careful reading of the manuscript.
\end{acknowledgments}

%%%%%%%%%%%%%%%%%%%%%%%%%%%%%%%%%%%%%%%%%%%%%%%%%%%%%%%%%%

%%%%%%%%%%%%%%%%%%%%%%%%%%%%%%%%%%%%%%%%%%%%%%%%%%%%%%%%%%

\end{document}